\documentclass[12pt,draft]{article}

\usepackage{amsthm,amssymb,amsmath,amscd,amstext,mathrsfs}
\usepackage{latexsym,mathptmx}

\newcommand{\nc}{\newcommand} 

\nc{\cb}{{\mathscr B}}
\nc{\cs}{{\mathscr S}}

\nc{\N}{{\mathbb N}}
\nc{\Z}{{\mathbb Z}}
\nc{\T}{{\mathbb T}}
\nc{\R}{{\mathbb R}}
\nc{\C}{{\mathbb C}}
\nc{\HH}{{\mathbb H}}

\nc{\dd}{{\rm d}}

\begin{document}

\title{Primordial black holes from collapsing antimatter} 

\author{G\'abor Etesi\\
\small{{\it Department of Geometry, Mathematical Institute, Faculty of
Science,}}\\
\small{{\it Budapest University of Technology and Economics,}}\\
\small{{\it Egry J. u. 1, H \'ep., H-1111 Budapest, Hungary}}
\footnote{E-mail: {\tt etesi@math.bme.hu}}}

\maketitle

\pagestyle{myheadings}
\markright{G. Etesi: Primordial black holes from collapsing antimatter}

\thispagestyle{empty}

\begin{abstract} 
In this paper a simple (i.e. free of fine-tuning, etc.) new mechanism for 
primordial black hole formation based on the collapse of large antimatter 
systems in the early Universe is introduced. A peculiarity of this 
process is that, compared to their material counterparts, the collapse of 
large antimatter systems takes much less time due to the reversed 
thermodynamics of antimatter, an idea which has been proposed in 
our earlier paper \cite{ete}.

This model has several testable predictions. The first is that the 
photon-baryon ratio is roughly computable and is equal to $1.95\times 
10^9$ which is quite close to its experimentally confirmed value. The 
second is that the mass of black holes arising from this mechanism is 
at least $10^5$-$10^6M_\odot$ hence they contribute to the super- or 
hypermassive end of the primordial black hole mass spectrum. The third 
prediction is that these sort of primordial black holes constitute at 
least $20\%$ of dark matter. Last but not least the observed current 
asymmetry of matter and antimatter, even if their presence in the 
Universe was symmetric in the beginning, acquires a natural 
explanation, too. 
\end{abstract}

\centerline{PACS numbers: 01.55.+b; 03.75.Hh; 04.70.-s; 05.70.-a}
\centerline{Keywords: {\it Primordial black holes; 
Matter-antimatter asymmetry; Second Law}}

%%%%%%%%%%%%%%%%%%%%%%%%%%%%%%%%%%%%%%%

\section{Introduction}
\label{one}

%%%%%%%%%%%%%%%%%%%%%%%%%%%%%%%%%%%%%%%%%

Phenomena of the physical world, as immediately given to us, appear in 
inexhaustable structures and formations of matter. At first sight a 
simple quantitative comprehension is achieved by understanding 
how much amount of matter a given fixed spatial region can accommodate. 
Approaching this way despite the endless possibilities one discovers two 
limits for matter formation: the lower 
universal limit is realized by an elementary particle (more precisely a 
relativistic quantum field) while the upper one is attained  
by a black hole; then one quickly arrives at the standard traditional and 
apparently disconnected territories of relativistic quantum field theory and 
the theory of gravity (general relativity). However this straightforward 
division into a linear and monotonic scheme extending 
from the ``smallest'' (which is something like an atomic thing) towards the 
``largest'' (which is something like a very different celestial thing) is 
too narrow. While the masses and sizes of elementary particles 
are indeed very small and are sharply restricted by yet unknown quantization 
rules such that the formers are below $m_{\rm Planck}\approx 
10^{-5}{\rm g}$ and the latters are above the corresponding Compton wave 
length $r_{\rm Planck}\approx 10^{-35}{\rm m}$, on the contrary black 
holes can in principle bear an arbitrary mass and size ranging from 
$m_{\rm Planck}$ with corresponding Schwarzschild radius $r_{\rm Planck}$ up to 
$1.2\times 10^{43}$ g and $1.8\times 10^{13}$ m (the data of 
the recently directly observed supermassive central black hole in the M87 giant 
elliptic galaxy) or even higher. Heavy black holes, whose existence has 
already been experimentally verified, indeed resemble astrophysical objects 
and have suitable origin however smaller-and-smaller black holes, if exist, 
exhibit more-and-more particlelike features; therefore the hypothetical 
borderline entity with mass $m_{\rm Planck}$ and size $r_{\rm Planck}$ can 
equally well be treated as either an extremely heavy particle or an extremely 
light black hole. Thus the apparently linear 
hierarchy of matter organization in Nature with its two limits rather 
would take a circular shape (if e.g. small black holes indeed exist).

A promising, even experimentaly confirmed candidate for a reservoir of 
small(er) black holes is cosmic dark matter. It is very likely a dark 
cocktail of various currently only hypothetical physical entities such 
as primordial black holes (PBHs) including evaporation remnants and yet 
mainly unknown weakly interacting massive particles (WIMPs) like 
neutrinos, axions, etc. The idea of a primordial black hole was 
introduced by {\it Hawking} 50 years ago \cite{haw2} and it was recognized soon 
\cite[p. 403]{car-haw} that the majority of matter might exist in 
the form of (primordial) black holes in the present Universe. There has been 
an intense debate recently among cosmologists and particle physicists 
concerning the ratio of the various dark matter candidates (it is impossible to 
give a complete list of references here therefore we refer here and from 
now on at other places to the excellent up-to-date review \cite{car-kuh} and 
the hundreds of references therein). Although we are still far from being 
conclusive according to diverse and accurate observations at least four 
mass windows are open for a primordial black hole abundance: these are 
the $10^{-16}$-$10^{-10}M_\odot$ together with the $10^{-6}$-$10^{-5}M_\odot$ 
windows in the small black hole range, the $10$-$10^3M_\odot$ window in 
the medium range and the larger than $10^{13}M_\odot$ spectrum in the 
hypermassive range, cf. \cite[Figure 1]{car-kuh}. It is not unreasonable 
that even our outer Solar System harbours a small black hole \cite{sch-unw}.

Sudden and violent primordial black hole formations during the course of 
the evolution of the Universe are usually associated with phase 
transitions of all kinds, cf. \cite{car-kuh} and in particular 
\cite{asa-gri-kuz-sha, car-cle-gar, gar-car-cle}. The general pattern is that 
the later the black hole formation occurs the higher the achieved black hole 
mass is \cite{car-haw}. In this paper a particular late-time phase 
transition, namely the photon recombination time around $380.000$ years 
after the Big Bang is examined from the point of view of massive 
primordial black hole formation. Our aim here is to offer a new mechanism 
based on a reversed thermodynamical behaviour of antimatter introduced 
in our earlier paper \cite{ete}. 

Already in 1939 {\it von Weizs\"acker} noted that the obvious 
but subjective difference between the past and future in our 
temporal experiences gains an objective substantiation by understanding 
the very content of the second law of thermodymanics \cite{weiz}. This 
understanding, among other consequences, would make the artificial division 
of time i.e. duration into a collection of disjoint and durationless instants, 
as motivated by the usual set-theoretic model of the continuum and 
assumed everywhere in physics, doubtful. While contemplating along these 
lines about the structure of time and its role played in current physical 
theories (for a survey cf. e.g. \cite{kum,sac}), the idea that macroscopic 
antimatter follows a {\it reversed} form of the second law of thermodynamics 
has been proposed \cite{ete}. Our suggestion is perhaps not independent of 
{\it Feynman}'s original ideas around 1947 that antiparticles should be 
regarded as ordinary (i.e. positive energy) particles but travelling 
backwards in time \cite{fey}. The {\bf Proposal} (see its discussion in 
Section \ref{two} below) implies that even if the {\it states} of a 
macroscopic matter and a macroscopic antimatter system are strictly identical 
on a ``snapshot'' taken at a fixed but purely hypothetically existing moment, 
their observable {\it temporal behaviour} 
is yet different and this difference is characterized by the usual and the 
reversed form of the second law. 
If one is indeed willing to accept that identical mechanical 
states might imply different temporal behaviour for matter and antimatter in 
their thermodynamical limit then one in fact questions a basic concept of 
Hamiltonian mechanics, namely the {\it state}. However the 
original Hamiltonian notion of a state which works well in traditional 
(i.e. e.g. antimatter-free) mechanics became already problematic in 
the 1930-40's (as {\it Feynman}'s idea also indicates) when physicists tried 
again to work out a model for the classical or the relativistic quantum field 
theoretic electron which is free of self-energy and other divergence 
problems.\footnote{As an aside 
we remark that the difficulty of assigning intermediate (i.e. non-asymptotic) 
states to interacting relativistic quantum fields was one of the 
theoretical---among other, including experimental---reasons 
why quantum field theory became a theory of scattering instead of the theory of 
states like traditional quantum mechanics \cite{blu}.} We do not intend to 
discuss here the deep problems arising from the division of the 
continuum into disjoint constituents \cite{bae}; 
rather point out that even if the {\bf Proposal} sounds weird it cannot be 
easily refuted by assuming its validity and then seeking a contradiction with 
some part of classical mechanics: for the {\bf Proposal} is the logical 
negation of the usual second law finding such a contradiction would be 
logically equivalent to a proof of the second law of thermodynamics from the 
laws of mechanics which is a very difficult (if not impossible) problem since 
{\it Boltzmann}'s times. In our opinion the validity of the {\bf Proposal} is 
an experimental question. 

Hopefully motivated with these introductory remarks in some extent, in 
this paper, in the realm of the structure of time we shall revisit the 
problem of the absence of antimatter from the Universe on macroscopic 
scales. What we are going to do is simple: instead of trying to derive 
the second law of thermodynamics from other abstract laws of theoretical 
physics we shall regard it as an underivable, irreducible, fundamental 
law expressing a basic {\it empirical evidence} about the temporal 
behaviour of macroscopic matter. The consequent application of 
considering the second law as an empirical evidence imposes at least one 
non-trivial constraint on its appearance in the physical world namely in 
its known form it is immediately applicable only to {\it ordinary} 
matter for this is the only form of matter which we have direct 
phenomenological contact with. Then we exhibit one plausible argument, 
based on various principles of theoretical physics 
but referring to the aforementioned observational validity of the 
second law, that the second law continues to hold for large antimatter 
systems but in a {\it reversed} form. Their converse thermodynamic 
behaviour could then lead to their swift confinement behind black hole 
event horizons hence to the absence of antimatter on marcroscopic scales 
from the Universe (also cf. \cite{boy-fin-tur, coh-kap, tou-tre-wil-zee}). 
Consequently the problem of missing antimatter \cite{coh-der-gla,ste1} 
naturally connects with the formation and frequency of black holes in the 
early Universe \cite{asa-gri-kuz-sha, car-cle-gar, gar-car-cle}. The 
mechanism we offer 
here sounds appealing for it does not require any fine-tuning or 
new asymmetric mechanism around Big Bang times to explain the macroscopic 
matter-antimatter asymmetry, as usually assumed in string theoretic and 
other approaches. 

The paper is organized as follows. In Section \ref{two} for completeness 
and the reader's convenience we recall from \cite{ete} 
the {\bf Proposal} but in a substantially improved form. Then in 
Section \ref{three} we apply it for the early Universe and introduce a new 
primordial black hole formation mechanism.

%%%%%%%%%%%%%%%%%%%%%%%%%%%%%%%%%%%%

\section{A proposal and its consequence}
\label{two}

%%%%%%%%%%%%%%%%%%%%%%%%%%%%%%%%%%%%%

The idea of an elementary antiparticle had quite unexpectedly dropped out 
from the theoretical efforts to reconcile the basic principles of special 
relativity and quantum mechanics; shortly thereafter their individual 
existence was verified by cosmic ray detectors, nuclear reactors and 
high energy particle colliders. However {\it no} physical experiment or 
even any kind of human experience in the broadest sense exists so far 
which could provide some phenomenological insight into the {\it 
macroscopic} i.e., {\it thermodynamical} properties of pure antimatter 
built up from bound states of these antiparticles. Even assuming that 
the basic principles of (classical or quantum) statistical mechanics 
continue to hold for physical systems consisting of pure antimatter---{\it 
and confessing that the derivation of the second law of thermodynamics 
from these principles is problematic yet}---the thermodynamical behaviour 
of such alien macroscopic physical systems is, honestly speaking, unknown 
to us presently. Therefore we are not in contradiction with any element of our 
contemporary description of physical reality if we make the following bit 
counterintuitive
\vspace{0.1in}

\noindent{\bf Proposal}. {\it Let $\cs_{\rm antimatter}$ be a closed 
physical system consisting of pure antimatter (in the low energy and 
thermodynamical limit). Then the entropy $S$ of this system never increases 
in time i.e., $\Delta S(\cs_{\rm antimatter})\leqq 0$.}
\vspace{0.1in}

\noindent In our opinion the ultimate validity or invalidity of the {\bf 
Proposal} is an experimental question; it can be surely decided by 
experiments designed to unfold the dynamics of large antimatter systems.

The property of being (anti)matter is Lorentz invariant i.e., it cannot be 
switched by Lorentz transformations. 
Therefore, as an immediate consistency check we note that the {\bf Proposal} 
is Lorentz invariant as well. This means that for any physical system 
$\cs$ (evolving forward in time) the sign of its entropy change, i.e. 
${\rm sign}(\Delta S (\cs))=\pm 1$ or $0$ in case of 
equilibrium, is invariant under Lorentz transformations despite that the 
entropy function $S(\cs)$ itself as usually defined in 
phenomenological thermodynamics or statistical mechanics is not obviously 
a Lorentz scalar. Indeed, let $\cs$ be a macroscopic physical system 
evolving along a future-directed non-spacelike congruence in 
Minkowski space-time and let $\gamma$ be a 
(co-moving or nearby, etc.) observer i.e. a future-directed timelike curve; 
define the {\it entropy change of $\cs$ 
with respect to $\gamma$} as the difference of the 
entropy of $\cs$ at a system-event observed as the {\it later} event 
$\gamma (\tau +\varepsilon)$ minus the entropy of $\cs$ at a system-event 
observed as the {\it earlier} event $\gamma (\tau)$ i.e., 
\[\Delta S(\cs,\gamma):=S(\cs,\gamma (\tau+\varepsilon )) 
-S(\cs,\gamma (\tau))\:\:.\] 
Let $\gamma'$ be another (perhaps distant) observer and 
define $\Delta S(\cs ,\gamma')$ analogously. Since $\cs$ evolves causally its 
two system events above are {\it not} spacelike separated consequently 
the observer $\gamma'$ records them in the same causal order: it 
observes the system-event corresponding to $\gamma (\tau +\varepsilon )$  
later than the system-event corresponding to $\gamma (\tau)$, too. 
Consequently even if perhaps 
$\Delta S (\cs,\gamma')\not=\Delta S(\cs,\gamma)$, we are sure that 
at least ${\rm sign}(\Delta S(\cs,\gamma'))={\rm sign}(\Delta S(\cs,\gamma))$ 
i.e. ${\rm sign}(\Delta S (\cs))$ is well-defined as stated.

Regarding its current experimental status, although 
as a hint for the {\bf Proposal} it is worth revisiting the already observed 
time asymmetry in various process governed by the weak interaction 
\cite{ang,ber-mar-vil,kab}, one has to acknowledge that we are still very 
far from a sharp experimental evaluation of the {\bf Proposal}. This is 
because despite the discovery of antimatter more than a half century ago 
only a very few types of antielements (namely $^{1}\overline{{\rm H}}$, 
$^{2}\overline{{\rm H}}$, $^{3}\overline{{\rm H}}$ and 
$^{3}\overline{{\rm He}}$, $^{4}\overline{{\rm He}}$) could have been produced 
so far and typically for very short times and in atomistic amounts only. 
However after taking an overview of these efforts we can select 
for our purposes the most relevant one namely the ALPHA experiment at CERN, 
which is a very exciting ongoing experiment exhibiting lot of new (but on 
theoretical grounds expected) facts about antihydrogen atoms, to see whether 
or not the communicated results can be used to support or reject the 
{\bf Proposal}. 
Latest results have been reported in \cite{bak} however from our point of 
view, i.e. regarding some technical background details, we shall revisit an 
older paper \cite{alp} from 2011 too. Recall that the original as well as 
present aim of the ALPHA together with the AEgIS experiment at CERN is to 
create and trap antihydrogen isotopes in order to carefully compare their 
physical properties with their ordinary counterparts. These physical 
properties are their lifetime (i.e. stability), 
spectrum, moreover soon gravitational characteristics like their mass and 
gravitational acceleration, too. In other words, and one should keep 
in mind this, the ALPHA and AEgIS experiments first of all have been 
technically designed to obtain precise information about properties of 
{\it individual} antiatoms. Nevertheless, since during the experiments 
thermal ensemble of antihydrogen atoms have regularly been produced, one 
expects to gain at least a marginal insight into their {\it collective} 
behaviour, too. 

The ALPHA experiment roughly goes as follows \cite[Figure 1]{alp} and 
\cite[Figure 1]{bak}. Using CERN's antiproton and positron accelerators 
and decelerators, soft antiproton and positron beams are injected into a 
tube of 280 mm axial length and of 44.35 mm diameter. The interior of this 
tube is vacuous and kept at low temperature, has optical access, and 
fulfilled with a strong magnetic field; it is actually a magnetic trap which is 
capable to confine those antihydrogen atoms which, after the recombination of 
the antiproton-positron plasma, can sufficiently rapidly cool down via 
advanced auxiliary laser cooling. Impressively, these atoms then can be 
trapped for several hours inside the vacuum tube to perform experiments. The 
vacuum tube is surrounded by silicon detectors to record final 
annihilations caused by interactions with the environment. This environment 
contains the tube's boundary, residual gases inside the vacuum tube as 
well as incoming particles from cosmic radiation and other accidental sources. 
Thus of course this environment consists of ordinary matter having 
standard thermodynamical properties. What from 
our viewpoint relevant is the values of the following three parameters in every 
{\it individual attempt} or run of the experiment: the number $N$ of 
trapped antihydrogen atoms, their temperature $T$ and their confinement 
time $t$. Although the cumulative value of $N$ was 
reported to be about 1000 in \cite{alp, bak}, its average 
value in individual attempts (i.e. the situation when antiatoms are 
under sharp observational control), as summarized in \cite[Table 1 and 
Figure 2]{alp} (but not available in \cite{bak}), was $N\approx 1$. 
Regarding the further parameters $T\approx 10$-$100$ mK and $t\approx$ several 
hours. We can now make three observations. Firstly, 
despite the vacuum tube's macroscopical volume 
$V\approx 3.95\times 10^{-5}$ ${\rm m}^3$ the entropy 
$S=S(N,T,V)$ of this antihydrogen gas system is practically zero (in accord 
with the third law of thermodynamics); consequently $S$ is practically constant 
despite the long observational time $t$ of any attempt. Secondly, 
the antihydrogen gas as observed in the ALPHA 
experiment cannot be considered as an ideal gas since its individual 
antihydrogen atoms are magnetically trapped generating strong 
correlations between their e.g. speed components (see e.g. 
\cite[Figure 3]{alp}). Therefore the truely free random motion of 
antihydrogen atoms 
in space, which is essential to study the temporal behaviour of their 
population's entropy, by design is not guaranteed even during long 
observational times. Thirdly and perhaps most importantly: the 
detectation of the position of an antihydrogen atom is based on its 
annihilation with the ordinary matter environment hence the position 
measurement procedure itself makes a strong ordinary thermodynamical 
influence on the antimatter system. (From our point of interest a more 
favourable position measurement protocoll should use e.g. low frequency i.e. 
soft photon scattering on the antihydrogen gas). To summarize: as, in our 
opinion, nobody could confirm the validity of the second law for 
normal hydrogen in a situation analogous to the ALPHA experiment, 
the current stage of this experiment is not suitable to challenge 
the {\bf Proposal} on an objective basis, too.

Regarding its current theoretical status, proving or disproving the {\bf 
Proposal} using the apparatus of theoretical physics and mathematics is 
at least as difficult as proving or disproving the ordinary second law. 
This is because the {\bf Proposal} is precisely the logical negation of 
the ordinary second law of thermodynamics; consequently a falsification 
of the {\bf Proposal} by assuming its validity and then arriving at a 
contradiction with some part of theoretical physics is logically 
equivalent to a proof of the second law (that is deriving it from the 
laws of classical or quantum statistical mechanics) by contradiction.

Having seen that challenging the {\bf Proposal} experimentally or theoretically 
is not straightforward, we would rather like to offer here one 
heuristic argument for its validity. In the following 
derivation of the {\bf Proposal} the validity of the second law as an 
{\it empirical evidence} about {\it macroscopic ordinary matter} systems will 
play a crucial role. This explains the absence of any kind of 
microscopic calculations from the considerations below: from the circle 
of our arguments it follows that the converse thermodynamical properties 
of antimatter is recognizable only macroscopically i.e. compared to that 
of ordinary particles, we are not going to modify the microscopic 
dynamics of antiparticles at all! Putting differently, 
one can say that the converse second law for antimatter is non-derivable from 
time-symmetric microscopic physics in exactly the same way as the ordinary 
second law is not derivable from it (yet).

{\it An argument based on the $CPT$ theorem of relativistic quantum field 
theories}. In light of our accurate experimental 
evidences, we have no reason to doubt the validity of the basic rules of 
relativistic quantum field theory when applied to both matter and antimatter. 
One of the most fundamental results of the relativistic quantum field theoretic 
description of physical reality is the $CPT$ theorem which states that 
the triple action of charge conjugation $C$, spatial reflection $P$ 
and time direction reversal $T$, when applied to a relativistic particle 
system, realizes a symmetry of it (cf. e.g. \cite[Chapter I.5.8]{wei2}). 
Since macroscopic matter is built up from the bound states of these 
relativistic particles it is reasonable to expect that the $CPT$ theorem 
continues to hold for low energy macroscopic physical systems in an 
appropriate {\it effective} form (for strongly related considerations cf. 
\cite{kli-maa}). We will assume two things: firstly 
that the physical system is built up from atomic (or molecular) matter in 
the {\it low energy thermodynamical limit} (this is certainly not true 
at the elementary particle level). This implies that the proposed $CPT$ 
violating mechanisms mainly based on various field oscillations between 
flavour-eigenstates (cf. e.g. \cite{ben-flo,cap-gia-lam,gag-san-tev-zuk,
guz-hol-oli,lis-mar-mon,sim-cap-gia}) are negligable in a good approximation: 
these $CPT$ violating effects are proportional to the mass difference between 
the flavour eigenstates of these free elementary particle fields but the 
occurence or the interaction with atomic matter of these states is 
negligable in the low energy limit. Secondly we assume that the parity 
transformation $P$ alone is already a symmetry of a physical system in the 
{\it low energy thermodynamical limit} (this is also not true at 
the elementary particle level). This assumption implies that performing $P$ 
on an existing low energy macroscopic physical system we obtain an existing 
low energy macroscopic physical system. 

Consider now an ordinary closed physical 
system $\cs_{\rm matter}$ consisting of pure (normal) matter in the low energy 
thermodynamical limit, evolving forward in time. Therefore, as 
a {\it theoretical consequence}, the $CPT$ theorem in its effective form  
is applicable to $\cs_{\rm matter}$ and tells us that 
\[CPT(\cs_{\rm matter}) =\cs_{\rm matter}\:\:\:.\] 
Another {\it empirical evidence} about $\cs_{\rm matter}$ is the validity of 
the second law of thermodynamics:
\[\Delta S(\cs_{\rm matter})\geqq 0\] 
i.e., the entropy of a closed physical system consisting of pure 
ordinary matter in the low energy thermodynamical limit 
never decreases. Putting together these we get 
\[\Delta S(CPT(\cs_{\rm matter}))\geqq 0\:\:\:.\] 
However, accepting the validity of the $CPT$ theorem in the low energy 
thermodynamical limit in an effective form discussed above, 
the $CPT$ transformation converts a closed physical system of matter evolving 
forward in time into a closed physical system containing (spatially reflected, 
hence existing) antimatter evolving backward in time i.e., 
\[CPT(\cs_{\rm matter})=\cs_{\rm antimatter\:\:in\:\:reversed\:\:time}\:\:\:.\] 
Therefore the last inequality implies  
\[\Delta S(\cs_{\rm antimatter\:\:in\:\:reversed\:\:time})\geqq 0\] 
i.e., the entropy of an antimatter system never decreases in reversed 
time hence switching back to ordinary time we come up with 
\[\Delta S(\cs_{\rm antimatter})\leqq 0\] 
leading to the {\bf Proposal}.

A comment on $CPT$ violation: recently there has been a debate 
concerning the (in)validity of the $CPT$ theorem on cosmological scales, 
in the presence of weak interaction, strong gravitational fields, 
etc., etc. (cf. e.g. \cite{ben-flo,boy-fin-tur,cap-gia-lam,
gag-san-tev-zuk,guz-hol-oli,kab,lis-mar-mon,sim-cap-gia}). Since our 
previous naive derivation of the {\bf Proposal} refers to the $CPT$ theorem 
and our considerations ahead deal with black holes in the early Universe 
it is worth addressing this issue here for a moment. 

Let $X,Y$ be some elementary particle states, 
denote by $P_{XY}(t)$ the probability of the occurence of the 
forward-in-time-process $X\rightarrow Y$ at a laboratory time $t$ and 
likewise $P_{YX}(t)$ the converse but also forward-in-time-process 
$Y\rightarrow X$. Introduce \cite{kab} the time-asymmetry parameter 
\[A_T(t):=\frac{P_{XY}(t)-P_{YX}(t)}
{P_{XY}(t)+P_{YX}(t)}\:\:.\]
If for example $X=K^0$ and $Y=\overline{K}^0$ are the neutral kaon and its 
antiparticle states then the {\it observed} violation of $CP$ in the kaon 
system together with the {\it theoretical assumption} of the validity of $CPT$ 
explains the {\it observed} $T$ violation i.e. $A_T\not=0$ in kaon experiments 
\cite{ang} (also cf. \cite{ber-mar-vil} for another example). However 
this temporal asymmetry already can be used alone to argue in favour to the 
{\bf Proposal} namely that the forward-in-time dynamics of kaon systems 
differs from that of their antiparticle counterparts (and this temporal 
asymmetry is generated by the weak interaction). Accordingly, most of the 
proposed $CPT$ violating mechanisms derive the violation itself from the {\it 
theoretical assumption} that in certain situations (e.g. large free particle 
systems in the presence of gravity \cite{sim-cap-gia}) $CP$ holds true but 
{\it finding theoretically} that in these situations $T$ fails because 
$A_T\not=0$. Therefore one may wonder whether or not in these situations the 
temporal asymmetry alone can be used directly (i.e. 
without referring to the $CPT$ theorem as we did before) to argue for 
(some form of) the {\bf Proposal}.

To close this section we discuss one consequence which plays a crucial role 
in our considerations ahead. We begin with 
clarifying that from now on by ``accepting the {\bf Proposal}'' in 
case of a macroscopic antimatter system $\cs_{\rm antimatter}$ we shall mean 
the following: this system obeys the same physical laws describing its physical 
{\it states} as its corresponding ordinary matter 
system $\cs_{\rm matter}$ defined by $\cs_{\rm matter}:= 
C(\cs_{\rm antimatter})$ where $C$ is the charge conjugation operator; 
however the physical laws describing the {\it dynamics} 
of $\cs_{\rm antimatter}$ might be different and are characterized by the 
{\bf Proposal} in an appropriate way. Then let us consider a closed 
macroscopical system $\cs_{\rm antimatter}$ built up from pure antimatter 
only hence not disturbed by recombination, etc. effects; thus the time 
evolution of $\cs_{\rm antimatter}$ is governed only by its own gravitational, 
electromagnetic and thermodynamical phenomena. Accepting the {\bf Proposal} 
therefore $\cs_{\rm antimatter}$ obeys the same equation of state 
(expressing a phenomenological relation between its energy, temperature, 
pressure, volume, etc.) as its corresponding macroscopic ordinary matter 
system $\cs_{\rm matter}$ however, unlike this latter, $\cs_{\rm antimatter}$ 
tends to evolve into more-and-more ordered states in time by its own dynamics. 
Since in case of $\cs_{\rm matter}$ 
the evolution into more-and-more disordered states often includes 
spatial expansion, the evolution of $\cs_{\rm antimatter}$ into 
more-and-more ordered states could imply its {\it stronger} tendency for 
spatial contraction. Consequently, in sharp contrast to an 
ordinary matter system, the structural tendency of $\cs_{\rm antimatter}$ for
spatial contraction in its own gravitational field could be {\it enhanced} by
the functional tendency of $\cs_{\rm antimatter}$ for spatial contraction
thanks to its reversed thermodynamics.\footnote{It is illustrive to
regard the structural and functional characters as sort of spatial and temporal
projections, respectively, of a common abstract ``character'' of a
physical system. In this language we can say that physical systems
possess an abstract ``contraction tendency'' whose structural and
functional manifestations are the gravity and the thermodynamical
phenomena, respectively (cf. Verlinde's idea of entropic gravity \cite{ver})
and they attenuate each other in the case of ordinary matter systems while
enhance each other in the case of antimatter systems.}

After these rather abstract general arguments let us examine the 
{\bf Proposal} and its consequences from a physically more realistic direction.

%%%%%%%%%%%%%%%%%%%%%%%%%%%%%%%%%%%%%%%%%%%%%%%%%%%%%

\section{An application to primordial black hole formation}
\label{three}

%%%%%%%%%%%%%%%%%%%%%%%%%%%%%%%%%%%%%%%%%%%%%%%

Consider the early Universe about the time when the last relevant, namely 
the electron-positron, spontaneous pair creation process stops because of 
cooling. Recall that in the radiation epoch $T\sim t^{-1/2}$ and for 
definiteness and simplicity we assume the temperature is about 
$T_0\approx10^9$ K and the 
time is about $t_0\approx10$ s after the Big Bang. Note that the physical 
description of the Universe already 
falls fully within the classical and non-(special)relativistic realm at these 
late times. Assuming the most natural initial conditions namely that the 
Universe was created with perfect particle-antiparticle symmetry and with 
precisely zero total electric charge we suppose that the Universe consists of 
an equal amount of baryonic-leptonic matter and antibaryonic-antileptonic 
antimatter of vanishing total electric charge surrounded with 
electromagnetic radiation, all in thermal equilibrium. Let us therefore 
model the whole situation with a 
closed classical thermodynamical system $\cs$ consisting of a finite 
spatial region fulfilled with a matter-photon-antimatter plasma in thermal 
equilibrium (whether or not $\cs$ can indeed be assumed to be closed will be 
addressed shortly). The expansion of the Universe is adiabatic 
hence its entropy is unchanged during (at least short times of the) expansion. 
We summarize all of these by writing symbolically 
$\cs=\cs_{\rm matter+radiation+antimatter}$ satisfying 
\begin{equation} 
\mbox{$\Delta S(\cs_{\rm matter\:\:+\:\:radiation\:\:+\:\:antimatter} )=0$ 
\:\: around \:\: $t_0\approx 10$ s.} 
\label{egyensuly1} 
\end{equation} 
We assume that the spatial region has volume $V_0$ which is much 
larger than the Debye length in the fulfilling plasma i.e. 
$V_0\gg\lambda^3_D\sim\left(\frac{T_0}{n_0}\right)^{3/2}$ where $n_0$ is 
the number density of the most dilute charged particle constituent in $V_0$ 
consequently the long-range electromagnetic interactions in the system are 
negligable because of screening-off. In a good approximation the only 
interaction between the matter-photon-antimatter subcomponents is 
annihilation or recombination of the various particle-antiparticle 
pairs. Since by our initial assumptions the electric charge of the observable 
Universe is zero at large temporal and spatial scales, as a single indicator 
for these various pair recombination processes we are going to monitor the 
main electron-positron recombination process, namely $e^-e^+\rightarrow2
\gamma$ only. However we note that this restriction, especially if 
neutrino effects are to be considered as well, can easily be relaxed in the 
following considerations if necessary.

Let us explore the time evolution of the system satisfying 
(\ref{egyensuly1}). Denoting by $t\gtrapprox t_0$ the time variable 
let $V(t)$ be the volume of $\cs_{\rm matter+radiation+antimatter}$ and 
$N(t)$ the number of electrons (or positrons) in it at a moment. More 
precisely let $N(t)$ be the {\it expectation value} at $t$ of the number of 
electrons in $V(t)$. Actually the true number of electrons 
in $V(t)$ essentially {\it never} coincides with the abstract number 
$N(t)$ rather takes its value somewhere in the interval $\left[\:N(t)- 
\sqrt{N(t)}\:,\:N(t)+\sqrt{N(t)}\:\right]$ due to thermal fluctuations 
i.e. accidentally entering and exiting particles. In other words 
strictly speaking our system is not closed. However if $N(t)$ is not 
the actual value but only the expectation value of the particle number, 
as we demand, then its very property is that it is independent of thermal 
fluctuations. Consequently with this definition of $N(t)$ the system 
$\cs_{\rm matter+radiation+antimatter}$ can indeed be assumed to be closed. 
These obvious but important remarks also imply that $N(t)$ depends on 
$t$ only through particle reactions which in our simple model means the 
single $e^-e^+\rightarrow2\gamma$ process alone.

Next let us therefore derive the evolution equation for this process. 
Of course a necessary condition for an electron-positron pair to 
annihilate in a fixed instant is that they should approach each other well in 
space; we capture this quantitatively by saying that 
if $\sigma (t)$ denotes the cross-section of the 
$e^-e^+\rightarrow2\gamma$ process then one particle must 
approximately stay within a ball of radius $\sqrt{\sigma(t)}$ about its 
antiparticle or {\it vice versa} during a short time interval $\Delta t$; 
consequently if $v(t)=\vert{\bf v}_{e^\pm}(t)\vert$ is the average speed of 
a particle and $\Delta t<\sqrt{\sigma (t)}/v(t)$ then the 
{\it effective annihilation volume} is not $V(t)$ but 
$v(t)\Delta t\sigma (t)N(t)$ only. Assuming uniform distribution the number 
of particles in this 
volume is $\big(v(t)\Delta t\sigma (t)N(t)/V(t)\big)N(t)$ which is 
therefore in a good approximation is equal to the number 
$N(t)-N(t+\Delta t)=-\Delta N (t)$ of annihilating pairs during $\Delta t$. 
Consequently letting $\Delta t\rightarrow 0$ the electron (or positron) 
number decreases according to 
\[\frac{\dd N(t)}{\dd t}=-\frac{v(t)\sigma(t)}{V(t)}N^2(t)\:\:.\]
The calculation of the cross-section of the 
$e^-e^+\rightarrow2\gamma$ process in the plane wave approximation 
(i.e. when the long range Coulomb forces are neglected) is a classical 
result of {\it Dirac}; since the plasma is already 
non-relativistic his quite complicated formula 
\cite[Equation (7) in Chapter V, \S27]{hei} reduces to its simple 
non-relativistic limit 
\[\sigma (t)\approx\pi r_0^2\:\frac{c}{2w(t)}\] 
where $r_0=e^2/m_ec^2\approx 2.82 
\times 10^{-15}$ m is the classical electron radius and $w(t)$ is the average 
speed of the colliding particles in their center-of-mass system hence
\[w(t)=\frac{1}{4\pi}\int\limits_{S^2}\left\vert{\bf v}_{e^\pm}(t)-
\frac{{\bf v}_{e^-}(t)+{\bf v}_{e^+}(t)}{2}\right\vert\dd\Omega=
\frac{1}{4\pi}\int\limits_{S^2}
\frac{\vert{\bf v}_{e^-}(t)-{\bf v}_{e^+}(t)\vert}{2}\dd\Omega =
\frac{2}{3}v(t)\:\:.\]
Note that the cross-section increases with time. 

Consider first the {\it radiation epoch} $10$ s $\lessapprox t\lessapprox
70.000$ a (here ``a'' stands for ``years'' as usual). 
Then $R (t)\sim \sqrt{t}$ implying $V(t)\sim t^{3/2}$. Therefore  
\[\left\{\begin{array}{lll} 
\dot{N}(t)&=&-\frac{3\pi r_0^2c t_0^{3/2}}{4V_0}t^{-3/2}N^2(t)\\
                  N(t_0) & = & N_0
         \end{array}\right.\]
where $V(t)=V_0\cdot (t/t_0)^{3/2}$ with $V_0>0$ being 
the initial volume at $t_0\approx 10$ s. Moreover 
$N_0>0$ is the initial particle number. The particular solution hence looks 
like 
\[N(t)=\left(\frac{1}{N_0}+\frac{3\pi r_0^2c t_0^{3/2}}{2V_0}
\left(\frac{1}{\sqrt{t_0}}-\frac{1}{\sqrt{t}}\right)\right)^{-1}\]
satisfying 
\begin{equation}
\lim\limits_{t\rightarrow +\infty}N(t)=\left(\frac{1}{N_0}+
\frac{3\pi r_0^2c t_0}{2V_0}\right)^{-1}
\label{aszimptotika1}
\end{equation} 
consequently having, quite surprisingly, a non-vanishing asymptotics thanks 
to the expansion.

Next, in the {\it matter epoch} i.e. when $70.000$ a 
$\lessapprox t\lessapprox 1.38\times 10^{10}$ a, then $R(t)\sim t^{2/3}$ 
yields $V(t)\sim t^2$. Therefore in the matter epoch 
\[\left\{\begin{array}{lll}
\dot{N}(t)&=&-\frac{3\pi r_0^2ct_1^2}{4V_1}t^{-2}N^2(t)\\
                  N(t_1) & = & N_1
         \end{array}\right.\]
where $t_1\approx 70.000$ a and $V_1=V_0\cdot (t_1/t_0)^{3/2}>0$ but now 
$V(t)=V_1\cdot (t/t_1)^2$. Moreover $N_1$ is the electron number at $t_1$. 
Note that by (\ref{aszimptotika1}) surely $N_1>0$ hence the corresponding 
matching particular solution in the matter epoch again looks like  
\[N(t)=\left(\frac{1}{N_1}+
\frac{3\pi r_0^2ct_1^2}{4V_1}
\left(\frac{1}{t_1}-\frac{1}{t}\right)\right)^{-1}\]
yielding 
\begin{equation}
\lim\limits_{t\rightarrow +\infty}N(t)=\left(\frac{1}{N_1}+
\frac{3\pi r_0^2ct_1}{4V_1}\right)^{-1}     
\label{aszimptotika2}
\end{equation}
hence has finite asymptotics, too. Note that without expansion i.e. putting 
$V(t)=$const. both solutions above would have trivial asymptotics 
$N(t)\sim t^{-1}$ i.e. the annihilation would be complete in this case.

Taking into account the electron-positron number asymptotics 
(\ref{aszimptotika1}) and (\ref{aszimptotika2}) together with the fact 
that the Universe is electrically neutral on large temporal and spatial 
scales, hence qualitatively all other particle (except probably the 
various neutrino) densities must follow more-or-less the same asymptotics, 
we end up with a rather surprising possibility: despite that their 
annihilation cross-section increases with passing time, in the 
sufficiently rapidly expanding Universe the matter and antimatter 
constituents do not annihilate completely. Although the previous 
considerations have been straightforward, the idea itself that antimatter 
could survive the early violent history of the Universe might look strange 
at first sight (although we note that various non-trivial freeze-out 
scenarios have already been studied by other authors, too cf. e.g. 
\cite{mur-sha-ski-koz, tho-dez-gro-kis}). Fortunately testable predictions 
derivable from this model help to measure the validity of this 
possibility. Perhaps the most directly accessable as well as measurable 
consequence is the photon-baryon ratio which is supposed to be somewhere 
between $10^9$ and $10^{10}$ in light of astronomical observations and 
cosmological considerations.

Thus let us make a digression here and see how this ratio looks like in our 
model. The initial number of
electrons is $N_0$ at the early moment $t_0\approx 10$ s meanwhile its late 
time limit is (\ref{aszimptotika2}) and their difference had been 
annihilated mainly into photons. Thus the quantity
$1/\eta_0:=\left(N_0-\lim\limits_{t\rightarrow+\infty}N(t)\right)\left/
\lim\limits_{t\rightarrow+\infty}N(t)\right.$ measures the magnitude of
the number of recombined electron-positron pairs hence the
photon-electron ratio. In its calculation just for simplicity the 
late time limit (\ref{aszimptotika1}) juxtaposed with (\ref{aszimptotika2}) 
can be replaced with the latter one alone if we write 
$t_0,N_0,V_0$ instead of $t_1,N_1,V_1$ in (\ref{aszimptotika2}). In this way 
we find
\[\frac{1}{\eta_0}=\frac{N_0}{\lim\limits_{t\rightarrow+\infty}N(t)}-1=
1+\frac{3\pi r_0^2ct_0N_0}{4V_0}-1=\frac{3\pi r_0^2ct_0\rho_0}{13k_BT_0}\]
where in the last step we inserted $\varepsilon_0=(3/2)k_BT_0$ and wrote
$\rho_0=(13/6)\varepsilon_0N_0/V_0$ for the baryonic matter energy 
density of the early Universe taking into account that the total $e^-+p^++n^0$ 
number in the initial volume $V_0$ was about 
$M_0\approx (1+1+1/6)N_0=(13/6)N_0$ at $t_0\approx 10$ s. Concerning the 
value of $\rho_0$ we assume (based on the measurements of $H_0$ 
and $\Omega_\Lambda$ in \cite{ade}) that the current total energy density 
of the Universe is $3.35$ GeV/${\rm m}^3$ having about $4\%$ 
baryonic constituent hence $\rho_0\approx 0.13 (T_0/2.71{\rm K})^3$ 
${\rm GeV}/{\rm m}^3$ with $T_0\approx10^9$ K. Plugging all the constants 
into the formula we obtain $1/\eta_0\approx 1.36\times 10^9$. From this number 
the photon-baryon ratio arises as follows. The total $e^-+p^++n^0$ number at 
late times is $M\approx (1+1+1/7)\lim\limits_{t\rightarrow+\infty}N(t)$ thus 
the total number of annihilating pairs is $(M_0-M)/M\approx 1/\eta_0$. Now let 
us do photon counting. The $e^-e^+\rightarrow 2\gamma$ process produces $2$
photons from every annihilating pair. In addition to this we have to consider
the leading low energy nucleon-antinucleon annihilations as well which are all 
the $p^+p^-$, $p^+\overline{n}^0$, $n^0p^-$ and 
$n^0\overline{n}^0\rightarrow X$ processes.
Referring to accurate particle collider results \cite{bac} the average number
of emitted photons in any of these pair recombinations is $\approx 3.93$.
Abandoning other decay channels (but for a broader survey of $N\overline{N}$ 
annihilations cf. \cite{kle-bat-ric}) these nucleon
processes together with $e^-e^+\rightarrow 2\gamma$ produce in
average $(3.93+ 2\times 3.93/7+3.93/7^2+2)/5\approx 1.43$ photons.
Consequently the photon-baryon ratio in our model looks like 
\[\frac{1}{\eta}\approx\frac{1.43}{\eta_0}\approx 1.95\times 10^9\]
which, taking into account the very rough estimates and simplifications we
made throughout,\footnote{Surely the most important of these 
simplifications was the systematic suppression of all neutrino effects 
including the decays $e^-e^+\rightarrow\nu_x\overline{\nu}_x$ 
where $x=e,\mu,\tau$. However as a consistency check note that since 
$\rho_0\sim T^3_0$ and $T_0\sim t^{-1/2}_0$ in the radiation era, at least 
the number $\eta_0=13k_B T_0/(3\pi r_0^2ct_0\rho_0)$ hence the baryon-photon 
ratio $\eta$ itself is independent of the particular choice $1$ s 
$\lessapprox t_0\lessapprox 10$ s for the initial value of time and 
the corresponding temperature $10^{10}$ K $\gtrapprox T_0\gtrapprox 
10^9$ K in the leptonic epoch we began with.\label{megj}} 
is close to its latest 
experimentally confirmed value $1/\eta\approx1.67\times 10^9$, cf. 
\cite{ade, ste2}.

Having seen that the recombination of matter with antimatter in the early 
Universe, such that the latter did not fully disappear from the stage 
during the course of this recombination era, produces a convincing 
photon-baryon ratio, one is unavoidably forced to say something on the 
following well-known fundamental problem formulated by {\it Sakharov} 
60 years ago \cite{sak1,sak2}: if this was indeed the story then what 
happened to antimatter \cite{sha}? Why 
large antimatter ``islands'' are not observable \cite{coh-der-gla, ste1} 
in the current Universe in spite of the obvious evidence 
that similar ordinary matter clouds do exist? This is the point where we evoke 
the {\bf Proposal}, which has not been used so far, i.e. apply it for large but 
isolated antimatter domains in the early Universe whose existence at 
least in the past follows from our previous considerations.

So let us continue the exploration of $\cs_{\rm matter+radiation+antimatter}$. 
As we have seen at late times it yet 
contains both matter and antimatter which essentially do not interact; 
consequently the original system splits into closed (or almost closed) 
subsystems what we write symbolically as 
\[\cs_{\rm matter+radiation+antimatter}=\cs_{\rm
matter}+\cs_{\rm radiation}+\cs_{\rm antimatter}\:\:.\] 
Hence by the (sub)additivity of the entropy the equilibrium equation 
(\ref{egyensuly1}) decouples as well consequently at for instance the hydrogen 
recombination time we re-write it as  
\begin{equation}
\mbox{$\Delta S(\cs_{\rm matter})+\Delta S(\cs_{\rm radiation})+
\Delta S (\cs_{\rm antimatter} )=0$\:\:around\:\:$t\approx 380.000$ a}\:\:.
\label{egyensuly2}
\end{equation}
Recall that in the matter epoch $T\sim t^{-2/3}$ more precisely at this moment 
$T\approx 3000$ K and the system contains neutral components 
(mainly photon gas and atomic hydrogen, helium 
together with their antimatter counterparts) only hence 
$\cs_{\rm matter}+\cs_{\rm radiation}+\cs_{\rm antimatter}$ can be treated well 
with the traditional tools of ideal gas theory in phenomenological 
thermodynamics from now on. The thermal equilibrium of the global system 
however does not necessarily implies the thermal equilibrium of its 
(weakly interacting) subsystems. This means that we have to examine them 
separately. Regarding $\cs_{\rm matter}$ an {\it empirical evidence} 
(i.e. not a theoretical deduction) about {\it this and only this} subsystem 
is again the validity of the second law
\begin{equation}
\Delta S(\cs_{\rm matter})\geqq 0\:\:.
\label{anyag}
\end{equation}
Within our closed system the next subsystem is $\cs_{\rm radiation}$ 
consisting of pure thermal radiation in equilibrium at temperature 
$T(\cs_{\rm radiation})$ occupying a volume 
$V(\cs_{\rm radiation})\lessapprox V(t)$. By the 
Stefan--Boltzmann law $S(\cs_{\rm radiation})=
\frac{4}{3}aT^3(\cs_{\rm radiation})\:V(\cs_{\rm radiation})$. But 
$T(\cs_{\rm radiation})\sim t^{-1/2}$ and 
$V(\cs_{\rm radiation})\sim t^{3/2}$ in the radiation era while   
$T(\cs_{\rm radiation})\sim t^{-2/3}$ and 
$V(\cs_{\rm radiation})\sim t^{2}$ in the matter era. Thus we find that   
\begin{equation}
\Delta S(\cs_{\rm radiation})=0
\label{sugarzas}
\end{equation}
which is of course in agreement with observations, too. 
(Note that this equality can be obtained via the {\bf Proposal} as well 
by applying it together with the usual second law on 
a closed physical system consisting of particles equal to their own 
antiparticles like $\cs_{\rm radiation}$; then indeed we can write both 
$\cs_{\rm radiation}=\cs_{\rm antimatter}$ and $\cs_{\rm radiation}=
\cs_{\rm matter}$ implying both $\Delta S(\cs_{\rm radiation})=
\Delta S(\cs_{\rm antimatter})\leqq 0$ and 
$\Delta S(\cs_{\rm radiation})=\Delta S(\cs_{\rm matter})\geqq 0$ hence 
(\ref{sugarzas}).) Regarding the last subsystem, namely 
$\cs_{\rm antimatter}$ we lack any observational support 
concerning its temporal behaviour however comparing (\ref{anyag}) and 
(\ref{sugarzas}) with (\ref{egyensuly2}) we can conclude that 
\begin{equation}
\Delta S(\cs_{\rm antimatter})\leqq 0
\label{antianyag}
\end{equation}
in accord with the {\bf Proposal} in the particular case of large isolated 
antimatter systems in the early but already low energy Universe. 

Geometrically the system 
$\cs_{\rm matter}+\cs_{\rm radiation}+\cs_{\rm antimatter}$ on 
a long range of macroscopic scales contains both matter and antimatter 
subregions of more-or-less equal volumes surrounded by electromagnetic 
radiation. To be more visual and taking into account the overall 
gravitational contraction as well, we can assume that within the finite volume 
$V(t)$ at a fixed time the spatial subregion occupied by 
$\cs_{\rm matter}$ is a (disjoint) union of finitely many $3$ dimensional 
balls of different radii, mainly proportional to the Jeans length under 
these conditions \cite{bat}, and likewise for 
$\cs_{\rm antimatter}$ such that the complementum of these balls is 
fulfilled with $\cs_{\rm radiation}$. Having set 
up this natural picture let us consider the further time evolution of 
the system when $t\gtrapprox 380.000$ a. The temporal evolutions of 
$\cs_{\rm matter}$ and $\cs_{\rm antimatter}$, which have been parallel up 
to this point, sharply split from now on. This is in some sense not surprising 
because $t\approx 380.000$ a is a crucial phase transition, namely the photon 
recombination, time in the history of the Universe.  

Regarding $\cs_{\rm matter}$ its individual ball constituents  
undergo standard star formation by gravitational contraction and further 
fragmentation. Recall that the first stage of this complex evolution is 
always an isothermal process in which a given ball radiates heat to its 
environment $\cs_{\rm radiation}$ during contraction. Consequently these 
balls are not closed systems taking into account this interaction.\footnote{But 
of course this radiative interaction is different from the one based on 
annihilation considered before but found to be already irrelevant at this 
asymptotic stage.} Nevertheless being of course an ordinary gas ball in 
a thermal bath observable, as an {\it empirical evidence} 
the second law surely applies to the in-this-way-interacting system  
$\cs_{\rm matter}+\cs_{\rm radiation}$ hence  
\[\Delta S(\cs_{\rm matter}+\cs_{\rm radiation})\geqq 0\:\:.\]
Moreover neither the volume change nor the companying thermal radiation of a 
gas ball in $\cs_{\rm matter}$ has effect on the volume or the temperature of 
its vast environment described by $\cs_{\rm radiation}$ hence in a good 
approximation $\Delta S(\cs_{\rm matter}+\cs_{\rm radiation})\approx 
\Delta S(\cs_{\rm matter})+\Delta S(\cs_{\rm radiation})$ thus 
via (\ref{anyag}) and (\ref{sugarzas}) we find that in fact 
$\Delta S(\cs_{\rm matter})\geqq 0$. That is, despite the radiative 
interaction, we can assume that $\cs_{\rm matter}$ alone satisfies the 
second law as usually assumed in standard textbooks on star formation 
\cite{zel}. 

After these preliminary observations take any particular ball within 
$\cs_{\rm matter}$ and treat it as a 
massive gas ball having volume $V$, particle number $N$, mass $m$, total 
energy $E$ and gravitational potential energy $U$ (hence kinetic energy 
$K=E-U$). Its entropy looks in the standard way like  
\[S(V,E)=Nk_B\left(\log\frac{V}{N}+
\frac{3}{2}\log\frac{E-U}{N}+{\rm const.}\right)\:\:.\] 
On substituting $V=(4\pi/3)R^3$ and the Newtonian potential energy 
$U=-Gm^2/R$ we can re-write it as 
\begin{eqnarray}
S(R,E)&=&Nk_B\left(3\log\frac{R}{N^{1/3}}+
\frac{3}{2}\log\frac{1}{N}\left(E+\frac{Gm^2}{R}\right)+{\rm const.}\right)
\nonumber\\
      &=& \frac{3}{2}Nk_B\left(\log\frac{ER^2+Gm^2R}{N^{5/3}}+
{\rm const.}\right)\nonumber
\end{eqnarray}
showing that $0<R<+\infty$ if $E\geqq 0$ or $0<R<-Gm^2/E$ if $E<0$. It 
turns out that the shape of the entropy function depends crucially 
on these two cases. Indeed, by solving the equation
\[\left.\frac{\partial S(R,E)}{\partial R}\right\vert_{E={\rm const.}}=
\frac{3}{2}Nk_B\frac{2ER+Gm^2}{ER^2+Gm^2R}=0\] 
it readily follows that if $E\geqq0$ i.e. the system is 
gravitationally not bounded then the entropy is a monotonly increasing 
function of $R$ hence without local extrema. Therefore taking 
any $0<R_{\rm initial}<+\infty$ and applying the second law for 
$\cs_{\rm matter}$ we find that $R\rightarrow +\infty$ hence the system 
is {\it unstable} in the usual sense: in order to maximize 
its entropy, a ball performs an unbounded spatial {\it expansion} as one would 
expect. The second possibility is that $E<0$ i.e. the system is gravitationally 
bounded then there is precisely one maximum of the entropy function at 
\[R=-\frac{Gm^2}{2E}\:\:.\] 
Therefore, depending on $R_{\rm initial}$ the system 
performs a finite {\it expansion}, remains {\it unchanged} or performs a 
finite {\it contraction} (by absorbing or releasing thermal radiation such 
that its temperature remains constant) in order to reach 
$R=-Gm^2/2E$ where $\cs_{\rm matter}$ attains its maximal entropy hence 
{\it stable} equilibrium state. This 
is the well-known isothermal phase of ordinary star formation.  

Now let us see how this analysis works for antimatter gas balls. 
Accepting the {\bf Proposal}\footnote{Cf. the end of Section \ref{two} what 
do we precisely understand by this.} $\cs_{\rm antimatter}$, which system in 
our astrophysical situation therefore describes a similar massive and in the
beginning cold pure antimatter ideal gas arranged into balls of
various radii, satisfies the usual equation of state $pV=Nk_BT$. 
Consequently taking one of its ball constituents the previous calculations 
work here as well. However this time we find a different temporal behaviour. 
The first possibility is when $E\geqq0$ hence this ball is gravitationally 
unbounded; then taking into account the reversed second law (\ref{antianyag}) 
we find that starting with any $R_{\rm initial}$ the radius of the ball 
behaves like $R\rightarrow 0$ that is, the antimatter gas ball is 
{\it unstable} again however in a reversed way: unlike an ordinary matter 
ball it performs an unbounded {\it contraction}. The second possibility is 
that the ball is gravitationally bounded i.e. $E<0$ and is 
already small i.e. $0<R_{\rm initial}<-Gm^2/2E$. Applying (\ref{antianyag}) 
we find again that $R\rightarrow 0$ i.e. a sufficiently small bounded 
antimatter gas ball is {\it unstable} in the sense that it undergoes an 
unbounded {\it contraction}, too. The third possibility is that the ball 
is gravitationally bounded i.e. $E<0$ and its 
volume is fine-tuned i.e. precisely $R_{\rm initial}=-Gm^2/2E$. Then this 
is an equilibrium of having maximal entropy hence is {\it unstable} in light of 
(\ref{antianyag}). The fourth possibility is that the ball 
is gravitationally bounded i.e. $E<0$ but is not small 
i.e. $-Gm^2/2E<R_{\rm initial}<-Gm^2/E$. Then by (\ref{antianyag}) it is 
{\it unstable} again but, quite surprisingly $R\rightarrow -Gm^2/E$ i.e. it 
would undergo a finite expansion such that the ball reaches a finite 
size. However the finite mass (and total energy) antimatter ball 
would have unbounded entropy in this limiting state which is a contradiction 
(if we regard the entropy as the logarithm of the number of microstates of a 
finite system). Hence the last two possibilities are ruled out which simply 
means that in our model the radius $R$, mass $m$ and total energy $E<0$ of a 
gravitationally bounded antimatter gas ball, when taking into account its 
dynamics i.e. temporal behaviour as well, always satisfies the inequality 
$R<-Gm^2/2E$. As a consistency check we remark that this maximal size is the 
half of the allowed maximal size for a gravitationally bounded ordinary 
matter ball of the same mass and negative total energy; this factor is 
reasonable in light of the stronger contraction tendency of antimatter: the 
idea we have stressed throughout the paper. Nevertheless, and to summarize, 
we have seen that in our model all of {\it these gravitationally unbounded or 
bounded antimatter balls suffer from a contractive instability}.

The characteristic collapsing time of this contractive instability 
can be estimated by the aid of the {\bf Proposal} alone (i.e. neglecting 
all other physical mechanisms for simplicity) if the interaction with 
the environment is also taken into account. Suppose 
$E_{\rm initial}<0$ and $0<R_{\rm initial}<-Gm^2/2E_{\rm intial}$ moreover 
that $\dot{E}<0$ hence $E<0$ throughout i.e. the gravitationally bounded 
ball radiates energy to its environment $\cs_{\rm radiation}$ during the 
contraction (cf. \cite[pp. 1197-1198]{kli-maa}) hence remains 
gravitationally bounded. Then putting $v:=\dot{R}$ applying 
(\ref{antianyag}) in the form 
\[\frac{\partial S}{\partial R}v+\frac{\partial S}{\partial E}\dot{E}
=\dot{S}\leqq 0\] and taking into account that $v<0$ we find that 
$\vert v\vert\geqq R^2\dot{E}/(2ER+Gm^2)>0$. If we further assume that 
$\dot{E}={\rm const.}$ then the estimated time required to contract from 
$R_{\rm initial}$ down to some $0<R<R_{\rm initial}$ is  
\[t_R=\int\limits_0^{t_R}\dd t=
\int\limits_R^{R_{\rm initial}}\frac{\dd r}{\vert v\vert}
\leqq\int\limits_R^{R_{\rm initial}}
\frac{2Er+Gm^2}{r^2\dot{E}}\dd r\leqq\frac{Gm^2}{\vert{\rm const.}\vert}
\int\limits_R^{R_{\rm initial}}\frac{\dd r}{r^2}=
\frac{Gm^2}{\vert{\rm const.}\vert}\left(\frac{1}{R}-\frac{1}{R_{\rm initial}}
\right)\]
which is finite even for the Schwarzschild radius 
$R=2Gm/c^2$ of the antihydrogen ball. 

Therefore taking into account their contractive instability discussed above 
one would expect that sufficiently massive antimatter gas clouds, 
compared to ordinary ones, are more capable to form black 
holes or enter already existing ones hence effectively feed 
them in very short times during the course of their 
dynamical evolution. Consequently, in light of the various uniqueness 
(``no-hair'') theorems of black hole physics (cf. e.g. \cite{heu}) pure 
macroscopic antimatter systems could disappear behind 
primordial black hole event horizons tracelessly faster in time than their 
ordinary counterparts. The details of why this antimaterial collapsing or 
feeding mechanism could be so effective are admittedly unclear at 
this stage of the art; perhaps the proposed reversed thermodynamics of 
antimatter somehow could prevent the system from friction hence the 
formation of high temperature radiating accretion discs, jets, etc. which are 
well-known refraining phenomena in case of observed compact objects 
swallowing normal matter. Although these important questions are open, 
for clarity we remark that the process itself is not in contradiction 
with Hawking's area theorem (cf. e.g. \cite{bar-car-haw,haw1}) because 
the fall of antimatter into a black hole, whatever weird its dynamical 
behaviour is, continues to transport further mass, electric charge and 
angular momentum into the black hole hence continues to increase the 
area of its instantaneous event horizon.\footnote{For example 
the horizon area 
$A=4\pi\left(2m^2-q^2+2m\sqrt{m^2-a^2-q^2}\:\right)$ of the Kerr--Newman 
black hole is invariant under $q\leftrightarrow -q$ i.e. the action of 
the charge conjugation operator $C$. Consequently from the point of view 
of black hole mechanics as summarized in \cite{bar-car-haw} it is inessential 
what sort of infalling thing, i.e. matter or antimatter, feeds the black hole. 
This is of course in agreement with the no-hair theorems.}

These qualitative considerations permit to make some testable predictions for 
primordial black hole physics. The first is that these black holes, due to 
their quite late born around the recombination time $t\gtrapprox 380.000$ a, 
are expected to be very massive. The mass of these black holes are 
related with the typical value of $R_{\rm initial}$, the radius of the 
collapsing ball examined above. Since the origin of these balls is the 
primordial fragmentation of the homogeneous but gravitationally unstable 
antihydrogen (and antihelium) gas, their radii are expected to satisfy $R_{\rm 
initial}\sim\lambda_J$ where $\lambda_J$ is the Jeans length around the 
recombination time. Hence the expected typical initial mass is proportional to 
the corresponding Jeans mass $m_J$ which by considerations still applicable 
here\footnote{Note that the standard computation of the Jeans length and mass 
rests only on mechanical i.e. reversible considerations.} is estimated 
in \cite{bat} to be $m_J\sim 10^5$-$10^6 M_\odot$. This large value is 
consistent with the general pattern that late-time born primordial black holes 
are expected to be heavier than the early-time ones. Consequently these black 
holes improve the less-understood super- or 
hypermassive end of the primordial black hole mass spectrum \cite{car-kuh}. 
Note that, on the contrary to their rapid formation, 
the typical high mass of these primordial black holes prevents them 
from too early evaporation by Hawking radiation \cite{haw3,wal}: since $t_{\rm 
evaporation}\approx 2.1\times 10^{67}(m_J/M_\odot)^3$ years which in this case 
is about $10^{82}$-$10^{85}$ years, these black holes do not reveal their 
content within our current cosmological times; consequently the details of 
their fate (i.e. the possible existence of evaporation remnants and their 
perhaps antimaterial nature, etc.) do not affect our considerations. The second 
prediction is that since the amount of matter is equal to antimatter and the 
latter had completely disappeared this way moreover the observed dark 
matter-baryonic matter ratio is about $5:1$ it follows that at least 
approximately 20$\%$ of the dark matter exists in the form of massive 
primordial black holes in our model. There has been recently an intense debate 
on the mass spectrum and the ratio of the primordial black hole constituent of 
dark matter; our predictions are consistent with current observational 
constraints as summarized in \cite{car-kuh}: regarding the mass spectrum 
observations permit the existence of an abundance of primordial black holes in 
the very heavy end of the mass spectrum while regarding the ratio all 
possibilities are apparently open (hence our 20$\%$ looks like a good 
compromise between the extremes).

%%%%%%%%%%%%%%%%%%%%%%%%%%%%%%%%%

\section{Conclusion}
\label{four}

%%%%%%%%%%%%%%%%%%%%%%%%%%%%%%%%%%%%%%%%%%%

In this paper two apparently independent problems of current cosmology: the 
basic problem of matter-antimatter asymmetry in the present Universe and 
late time primordial black hole formation has been connected by a proposed 
reversed thermodynamical behaviour of antimatter. Within this framework the 
observed baryon-photon ratio has been reproduced 
($\eta\approx 5.12\times 10^{-10}$ is our prediction) whose accuracy 
is convincing ($\eta\approx 5.99\times 10^{-10}$ is
the experimental value) taking into account the simplicity and naturality of 
its derivation carried out here. Moreover two testable predictions 
concerning the average black hole masses (which is larger than 
$10^5$-$10^6M_\odot$) and the primordial black hole ratio in dark matter 
(which is at least 20\% ) in this model has been exhibited. However 
the model's most appealing feature is surely a natural, 
effortless (i.e. free of any fine-tuning, etc.) explanation of the 
problem of missing antimatter. 

To close we emphasize once more that the idea proposed here requires further 
elaboration and we also admit that all of these rough qualitative 
considerations might be invalidated by exploring the highly complex details 
of time evolution of realistic physical systems including the effect of gravity 
on the $CPT$ theorem \cite{sim-cap-gia}; however these 
certainly very difficult analyses are beyond the 
limits of this short note. Nevertheless our considerations, perhaps together 
with other suggested mechanisms (far from being complete cf. e.g. 
\cite{arn-mcl, asa-gri-kuz-sha, boy-fin-tur, buc-plu,
car-cle-gar,coh-kap,far-sha, gar-car-cle,gar-gri-kus-sha, joy-pro-tur,
mat-dol-nag-sat,tou-tre-wil-zee,wei1}), might shed a light onto the 
origin of the observed matter-antimatter asymmetry in the current Universe, 
even if matter and antimatter was produced in symmetric amounts in the Big 
Bang. This asymmetric mechanism 
together with the symmetric recombination effects could be responsible for 
the deficit of antimatter as well as for the rapid early galaxy 
formation around supermassive primordial black hole cores in the 
observable Universe. 
\vspace{0.1in}

\noindent{\bf Acknowledgement.} The author is grateful to K. Bozsonyi, 
K. Furgason, L. Krasznahorkai and F. Sikl\'er for the stimulating discussions. 
There are no conflicts of interest to declare that are relevant to the content 
of this article. All the not-referenced contents in this work are fully 
the author's own contribution. No funds, grants, or other financial 
supports were received. The work meets all ethical standards applicable 
here.

\end{document}